\documentclass[conference]{IEEEtran}

\usepackage{cite}
\usepackage{amsmath,amssymb,amsfonts}
\usepackage{algorithmic}
\usepackage{graphicx}
\usepackage{textcomp}
\usepackage{xcolor}

\def\BibTeX{{\rm B\kern-.05em{\sc i\kern-.025em b}\kern-.08em
    T\kern-.1667em\lower.7ex\hbox{E}\kern-.125emX}}

\usepackage[caption=false,font=footnotesize]{subfig}
\usepackage{stfloats}
\usepackage{multirow}

\begin{document}

\title{Identifying Stable Influencers:\\
Distinguishing Stable and Temporal Influencers\\
Using Long-Term Twitter Data}

\author{\IEEEauthorblockN{Harutaka Yamada}
\IEEEauthorblockA{Graduate School of Science and
Technology\\
University of Tsukuba\\
Ibaraki, Japan 305--8573\\
Email: hrtk.ymd.2002@snlab.cs.tsukuba.ac.jp}
\and
\IEEEauthorblockN{Sho Tsugawa}
\IEEEauthorblockA{Institute of Systems and\\
Information Engineering\\
University of Tsukuba\\
Ibaraki, Japan 305--8573\\
Email: s-tugawa@cs.tsukuba.ac.jp}
\and
\IEEEauthorblockN{Mitsuo Yoshida}
\IEEEauthorblockA{Institute of Business Sciences\\
    University of Tsukuba\\
    Tokyo, Japan 112--0012\\
    Email: mitsuo@gssm.otsuka.tsukuba.ac.jp}
}

\maketitle

\begin{abstract}
For effective social media marketing, identifying stable influencers—those who sustain their influence over an extended period—is more valuable than focusing on users who are influential only temporarily.
This study addresses the challenge of distinguishing stable influencers from transient ones among users who are influential at a given point in time. We particularly focus on two distinct types of influencers: source spreaders, who widely disseminate their own content, and brokers, who play a key role in propagating information originating from others.
Using six months of retweet data from approximately 19,000 Twitter users, we analyze the characteristics of stable influencers. Our findings reveal that users who have maintained influence in the past are more likely to continue doing so in the future.
Furthermore, we develop classification models to predict stable influencers among temporarily influential users, achieving an AUC of approximately 0.89 for source spreaders and 0.81 for brokers.
Our experimental results highlight that current influence is a critical factor in classifying influencers, while past influence also significantly contributes, particularly for source spreaders.

\end{abstract}

\begin{IEEEkeywords}
Influencer, Social Network, Social Media
\end{IEEEkeywords}

\section{Introduction}
\label{sec:intro}

In recent years, companies have increasingly adopted social media as a key tool for branding and marketing. Social media is not only a communication platform but also an important source of information, especially among teenagers and students \cite{whiting2013people,aillerie2018social,arafah2023digital}. As a result, many brands, corporations, and organizations actively use social media platforms to disseminate information.

One prominent form of social media-based marketing is {\em influencer marketing}. Influencer marketing is a form of marketing in which companies select influencers, provide them with incentives, and have them promote the companies’ products or services on social media platforms\cite{leung2022online,campbell2020more}.

\begin{figure}[tbp]
\centering
\includegraphics[width=0.9\linewidth]{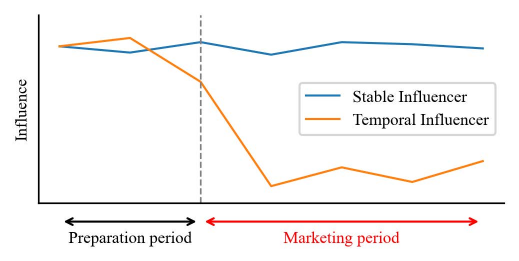}
\caption{Illustration of how influence evolves over time for a stable influencer and a temporal influencer.}
\label{fig:ex}
\end{figure}

To ensure broad and effective dissemination of content, it is crucial for companies to select stable influencers.
As illustrated in Fig.~\ref{fig:ex}, influencers with similar influence levels at a given time can differ significantly in their long-term effectiveness: some maintain influence over an extended period ({\em stable influencers}), while others enjoy only short-lived popularity ({\em temporal influencers}) \cite{Uehara}.
When entering long-term contracts with influencers, or even considering the time lag between initial outreach and campaign launch, it is essential to distinguish between these two types. Thus, identifying effective influencers requires a long-term perspective that considers how their influence evolves over time.

Despite its practical importance, the temporal dynamics of influence in social media remain an underexplored research area. Most existing studies identify influencers without distinguishing between stable and temporal influencers, often focusing on short timeframes or specific points in time. Research utilizing information diffusion histories is typically constrained by limited data, with observation periods usually spanning only one to two months or covering a single topic of discussion \cite{bakshy2011everyone,tsugawa2023identifying}.

In this paper, we analyze the long-term influence dynamics of influencers, and aim to clarify a method for identifying stable influencers among those who appear influential at a given point in time. In particular, we address the
 following research questions.
    \begin{itemize}
    \item \textbf{(RQ1)} What are the differences in characteristics between stable and temporal influencers?
    \item \textbf{(RQ2)} To what extent can we accurately predict whether an influencer is stable or temporal?
    \item \textbf{(RQ3)} If prediction is feasible, which features contribute to the prediction?
    \end{itemize}

To address these questions, we analyze fluctuations in user influence over a six-month period using tweet and retweet histories from 19,000 Twitter (currently X)
users. Based on changes in influence over time, we classify users as either stable or temporal influencers. We then examine their structural and behavioral characteristics within the social graph (RQ1). Using these features, we build classification models to distinguish between the two types of influencers and evaluate their prediction accuracy (RQ2). Finally, we analyze the feature importance in our models to identify key characteristics that contribute to identifying stable influencers (RQ3).

We also consider two distinct types of influencers on social media: source spreaders, who amplify the reach of their own content, and brokers, who spread content originally posted by others \cite{BURT2000345,Araujo04052017,tsugawa2023identifying}. Both play crucial roles in large-scale information diffusion \cite{pei2014searching,Araujo04052017}, and our study covers both.

Our main contributions can be summarized as follows.
    \begin{itemize}
    \item \textbf{Characterization of stable vs. temporal influencers:}\\
    We provide a detailed analysis of both stable and temporal influencers. Our results show that only about half of the users maintained their influence over a six-month period, regardless of whether they were source spreaders or brokers. Furthermore, our findings support the observation by previous work \cite{tsugawa2023identifying} that brokers cannot be effectively characterized using a single centrality measure; this holds true for both stable and temporal brokers.
    \item \textbf{Classification of influencer types:}\\
    We developed classification models to predict whether a given influencer is stable or temporal. For both source spreaders and brokers, the current influence score was a strong predictor. Notably, for brokers, features such as centrality measures were shown to be ineffective for prediction.
    \end{itemize}
    
The remainder of this paper is organized as follows.
Section~\ref{sec:related} reviews related work on influencer identification. Section~\ref{sec:method} defines an influencer and describes the experimental methodology. Section~\ref{sec:result} presents the results and discussion. Finally, Section~\ref{sec:conc} summarizes our findings and outlines directions for future work.

\section{Related Work}
\label{sec:related}

Various definitions of influencers have been proposed in the literature. Broadly speaking, an influencer on social media is a user who exerts significant influence over a large number of others. Influencers are often discussed in the context of information diffusion, and are typically defined as powerful spreaders capable of triggering large-scale dissemination through their own content or posts. In this paper, following \cite{tsugawa2023identifying}, we refer to such users as influential source spreaders.
In contrast, some studies have focused on brokers who can disseminate information originated by others to a large audience \cite{BURT2000345, Araujo04052017, tsugawa2023identifying}. Influential brokers serve as bridges across communities and are also essential to large-scale information diffusion \cite{Araujo04052017, LI20145115}.
In this paper, following \cite{tsugawa2023identifying}, we refer to such users as influential brokers.
Accordingly, in this paper, we adopt the classification of influencers into two types: source spreaders and brokers, and analyze them separately.

The task of identifying influencers in large-scale diffusion processes is a core problem in network science. A widely used approach involves computing centrality measures based on the underlying network structure. These include purely structural metrics such as degree centrality \cite{freeman1978centrality}, as well as iterative algorithms like PageRank \cite{LU20161}, which simulate random walks to capture more global structural properties. By applying such centrality measures to social networks, users with high centrality scores are typically identified as influential nodes.  Some studies combine multiple centrality features and apply machine learning techniques to improve influencer identification \cite{bucur2020top}. Another prominent line of research frames the problem as an instance of the Influence Maximization Problem \cite{kempe2003maximizing}, which seeks an optimal set of seed nodes that maximizes the spread of information in a network. This approach relies on diffusion models such as the linear threshold model and the independent cascade model, where simulations are used to estimate influence propagation. Additionally, several methods identify influential users based on their past diffusion performance, using historical activity data such as retweets and likes to estimate individual influence levels \cite{bhowmick2019temporal}.

The effectiveness of influencer identification methods has been evaluated using both simulated and real-world data. Simulation-based evaluations, often leveraging diffusion models \cite{huang2020identifying, kempe2003maximizing}, assess how much information spreads when selected influencers are used as seeds. In contrast, empirical evaluations use actual diffusion histories \cite{bhowmick2019temporal, Pei2018} to measure the extent to which identified users contributed to observed diffusion outcomes during a specific time period.
This study also utilizes historical diffusion data. While previous research using such data has generally been constrained to short observation period, we use six-months data of Twitter users for distinguishing stable and temporal influencers.

A few studies have adopted a longer-term perspective. For instance, Cha et al. \cite{cha2010measuring} examined retweet counts over an eight-month period for users identified using centrality measures, while Uehara and Tsugawa \cite{Uehara} analyzed how the influence of central users fluctuated over the course of a year. However, these existing studies focused on the persistence of influence and did not distinguish between stable and temporal influencers. Building upon these studies, our study advances the understanding of influencer dynamics by explicitly classifying users into stable and temporal influencers and analyzing their distinguishing characteristics.  Furthermore, unlike previous studies, we separately examine both source spreaders and brokers, providing a more comprehensive view of influence dynamics on social media.

\section{Methodology}
\label{sec:method}
  
\subsection{Dataset}

We constructed a Twitter dataset containing both user follow relationships and retweet activity. Monthly snapshots of follow relationships have been continuously collected for one million users randomly sampled from Twitter’s sample stream. In parallel, we have been collecting English-language retweets on an ongoing basis using the Twitter Search API. These two data sources serve as the foundation of our analysis.
The combined dataset covers English retweets and user follow relationships spanning from October 2021 to December 2022. To ensure longitudinal consistency and user activity, we restrict our analysis to users who (i) had at least one of their tweets retweeted in each month from January to December 2022, and (ii) had at least one follower during this period.
Table~\ref{tab:dataset} summarizes the total number of followers and retweets for all users in the final dataset as of January 2022.

\begin{table}[tbp]
\renewcommand{\arraystretch}{1.1}
\caption{Summary Statistics of the Dataset as of January 2022}
\begin{center}
\begin{tabular}{l|r}
\hline
\textbf{Metric} & \textbf{Count} \\
\hline
Num. users & 18,950 \\
Total num. followers & 63,656,198 \\
Total num. retweets & 12,620,483 \\
\hline
\end{tabular}
\label{tab:dataset}
\end{center}
\end{table}

\subsection{Network Construction and Notation}

To examine the characteristics of users, we constructed monthly networks representing follow and retweet (RT) relationships among users from the dataset. In each network, a node $v$ represents a user, and a directed link $(u, v)$ indicates that user $u$ either follows user $v$ or retweeted $v$'s post at least once during the given period. We denote the follower network for month $m$ as $G_m^{\rm FO}$ and the RT network as $G_m^{\rm RT}$. All networks are directed and unweighted.

As we will explain in the next subsection, we use the sequence of RTs for evaluating influence of users. Here, we define basic notations. Let the set of tweets posted by user $u$ be denoted as 
$P_u = \{p^u_1, p^u_2, p^u_3, \ldots, p^u_n\}$.
For a given tweet $p$ and a time window $\tau$, we define the set of retweets as 
$R^p_\tau = \{r^p_{u1}, r^p_{u2}, r^p_{u3}, \ldots, r^p_{un}\}$,
and their corresponding timestamps as 
$T^p_\tau = \{t^p_{u1}, t^p_{u2}, t^p_{u3}, \ldots, t^p_{un}\}$.
We denote the set of users who retweeted $p$ during $\tau$ as 
$U^p_\tau = \{u^p_1, u^p_2, u^p_3, \ldots, u^p_{n'}\}$.
Furthermore, for a given user $v$ who retweeted $p$ during $\tau$, we define the set of subsequent retweets as 
$D^{p,u}_\tau = \{r^p_n \mid t^p_n > t^p_u\}$.

\subsection{Definition of Source Spreaders and Brokers}

In this study, we define each user's source spreader score and broker score based on RT data from the dataset, and use these scores to identify source spreaders and brokers.

While there are various definitions of user influence, following previous studies on influencer detection \cite{tsugawa2023identifying, ye2010measuring}, we define a user's influence as the volume of reactions from other users to their posts. To quantify such reactions, we use the number of retweets a user's posts receive.
A user's source spreader score $S_{\tau}$ for month $\tau$ is defined as the total number of retweets of their tweets during that month, as follows:
\begin{equation}
\label{spscore}
S^{u}_{\tau} = \sum_{p \in P_{u}} |R^{p}_{\tau}|
\end{equation}

Similarly, a user's broker score $B_{\tau}$ for month $\tau$ is defined as the total number of subsequent retweets that occurred after the user $v$ retweeted a given tweet $p$ during that month:
\begin{equation}
\label{brscore}
B^{u}_{\tau} = \sum_{p} |D^{p,u}_{\tau}|
\end{equation}

For each month, users whose source spreader or broker scores fall within the top 10\% are identified as source spreaders or brokers, respectively.

We define \textit{stable} influencers as users who are identified as source spreaders or brokers for $m$ consecutive months. Those who do not meet this criterion are classified as \textit{temporal} influencers.

\subsection{Features of Users}

To investigate the characteristics of stable influencers, temporal influencers, and other users, we calculated their network features of each month.
From the networks constructed from the dataset, namely the follower network $G_m^{\rm FO}$ and the RT network $G_m^{\rm RT}$, we calculated in-degree and PageRank~\cite{PageRank} of each user, and community size, to which the user belongs. 
The in-degree in $G_m^{\rm FO}$ corresponds to the number of followers, while in $G_m^{\rm RT}$ it corresponds to the number of unique retweeters (RTer counts).  
For obtaining communities, we used the Leiden algorithm \cite{Leiden}.

We also calculated the features regarding the dynamics of influence of each user.  Here, we used the change rate of influencer score of a user from period $\tau$ to $\tau^{\prime}$.  The change rate $C^{u}_{\tau\to\tau^{\prime}}$ of user $u$ is defined as:
\begin{equation}
\label{change}
C^{u}_{\tau\to\tau^{\prime}} = \log {\frac{I_{\tau^{\prime}}}{I_{\tau}}}
\end{equation}
where $I^{u}_{\tau}$ is an influence score (either source spreader score or broker score) of user $u$ during a period ${\tau}$.

To quantify the diversity of users who retweet a give user's post, we calculated the unique user rate.
For a period $\tau$, the unique user rate $Q^{u}_{\tau}$ is defined as the ratio of the total number of unique retweeters $\left|\bigcup_{p\in P_u} U^p_{\tau}\right|$ across all posts $P_u$ by the user to the total RT counts $S^{u}_{\tau}$ during that period:
\begin{equation}
Q^{u}_{\tau} = \frac{\left|\bigcup_{p\in P_u} U^p_{\tau}\right|}{S^{u}_{\tau}}
\label{Uniq}
\end{equation}

Table~\ref{tabF} summarizes the features used in our analysis, along with their dimensionality and category labels. In the subsequent analysis, we use January 2022 as the reference month and examine user features over the four months leading up to and including this month. Accordingly, most features are represented as four-dimensional vectors, corresponding to these four months.
An exception is the change rate, which represents the rate of change in the influencer score between the first and second halves of the reference month and is therefore one-dimensional.
Category labels indicate whether each feature is derived from the follow network, calculated from retweet data, or based on the broker score.
The change rate is defined in two forms—one based on the broker score and another based on the source spreader score—so it is associated with both the RT and BR categories.

    \begin{table}[tbp]
    \caption{Features used in analysis}
    \begin{center}
    \begin{tabular}{l|r|c}
    \hline
    \textbf{Feature} & \textbf{Dimensions}& \textbf{Category} \\
    \hline
    Follower counts & 4 & Follow\\
    PageRank (Follower net)& 4 & Follow\\
    Community size (Follower net)& 4 & Follow\\
    RT counts & 4 & RT\\
    RTer counts & 4 & RT\\
    PageRank (RT net)& 4 & RT\\
    Community size (RT net)& 4 & RT\\
    Unique user rate & 4 & RT\\
    Influencer score change rate& \multirow{2}{*}{1} & \multirow{2}{*}{RT/BR}\\
    (First-to-second half of the reference month)&&\\
    Broker score & 4 & BR\\
    \hline
    \end{tabular}
    \label{tabF}
    \end{center}
    \end{table}

\subsection{Prediction Model for Source Spreaders and Brokers}

Our experiments aim to predict whether influencers identified in a specific month are stable or temporal influencers. We used January 2022 as the reference month and constructed training data based on whether spreaders and brokers identified in January remain influencers over subsequent months.
Each feature spans the four-month period from October 2021 to January 2022, as summarized in Tab.~\ref{tabF}. The training labels indicate whether a user continued to be an influencer for m consecutive months starting from January. Unless otherwise specified, we set $m=6$, meaning users must have remained influencers continuously from January through June.

We employ LightGBM \cite{lightgbm} to build the prediction model, using the features shown in Tab.~\ref{tabF}. The training and test datasets are split in a 7:3 ratio. Hyperparameters are tuned by performing 5-fold cross-validation and grid search on the training data.

For evaluation, we use data centered on July 2022. Similarly, features from July and the three preceding months (April 2022 to June 2022) are used, and the ground-truth labels indicate whether a user remained an influencer for $m$ consecutive months starting from July.

Prediction performance is evaluated using accuracy and AUC metrics. Accuracy measures the proportion of correctly classified instances, while AUC quantifies the area under the ROC curve, which plots the true positive rate (TPR) against the false positive rate (FPR) at various classification thresholds.

\section{Results and Discussion}
\label{sec:result}

\subsection{Long-Term Retention of Influence Among Social Media Users}

\begin{figure*}[!t]
\centering
\subfloat[Source spreaders]{
    \includegraphics[width=0.45\linewidth]{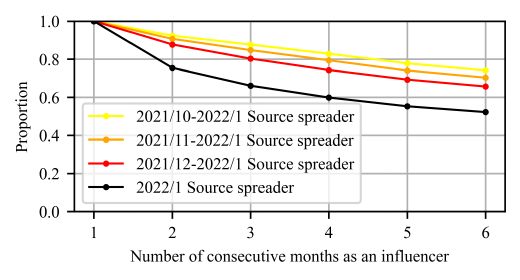}
    \label{figsp1}
    }
\hfil
\subfloat[Brokers]{
    \includegraphics[width=0.45\linewidth]{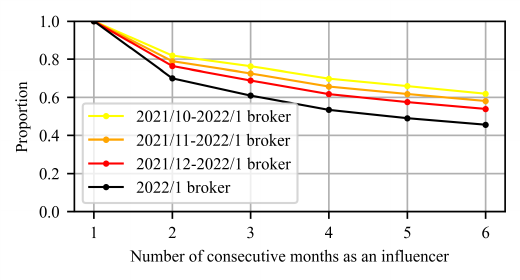}
    \label{figbr1}
    }
\caption{Proportion of users remaining influencers for consecutive months. (a) 52.2\% of users were stable source spreaders for six months. Among users who were identified as influential source spreaders from October to January, 74.1\% remained stable for the next six months. (b) 45.5\% of users were stable brokers for six months. 
Among users who were identified as influential broker from October to January, 61.9\% remained stable for the next six months.}
\label{figsp1figbr1}
\end{figure*}

We begin by analyzing how long users identified as influencers in a given month continue to retain their influence over time.
Specifically, we track users recognized as influencers in January 2022 and measure the number of consecutive months they continue to hold this status.
Additionally, we examine users who had already been identified as influencers for two months (December 2021--January 2022), three months (November 2021--January 2022), and four months (October 2021--January 2022) to assess how their influence persists beyond January.
Figure~\ref{figsp1figbr1} shows the relationship between the number of consecutive months a user is identified as an influencer and the proportion of such users, for both source spreaders and brokers.

From Fig.~\ref{figsp1figbr1}, we observe that approximately 52.2\% of users identified as source spreaders in January 2022 remained influencers for the subsequent six months. Similarly, around 45.5\% of users identified as brokers in January continued to be recognized as brokers for six consecutive months. These results indicate that even when users are identified as source spreaders or brokers based on a single month's retweet activity, nearly half of them fail to sustain their influence over the longer term.

Figure~\ref{figsp1figbr1} also reveals a clear trend: users with a longer history of influence are more likely to remain influential. Among those identified as source spreaders since December 2021, 65.6\% continued to hold that status for six months, compared with 70.3\% for those identified since November and 74.1\% for those since October.
A similar trend is observed for brokers: 53.9\% of those identified since December remained brokers for six months, increasing to 58.0\% for those since November, and 61.9\% for those since October. These results suggest that prior consistency in influence strongly correlates with the likelihood of continued influence.

\subsection{Analysis of Characteristics of Stable Influencers}

\begin{figure*}[!t]
\centering
\subfloat[Source spreaders]{
    \includegraphics[width=0.4\linewidth]{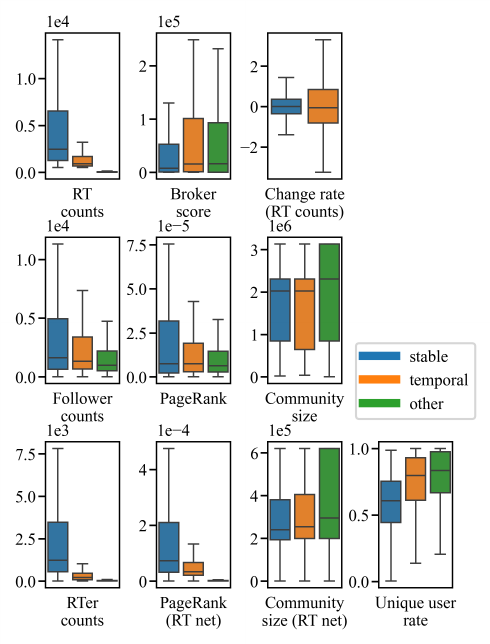}
    \label{fig1-2}
    }
\hfil
\subfloat[Brokers]{
    \includegraphics[width=0.4\linewidth]{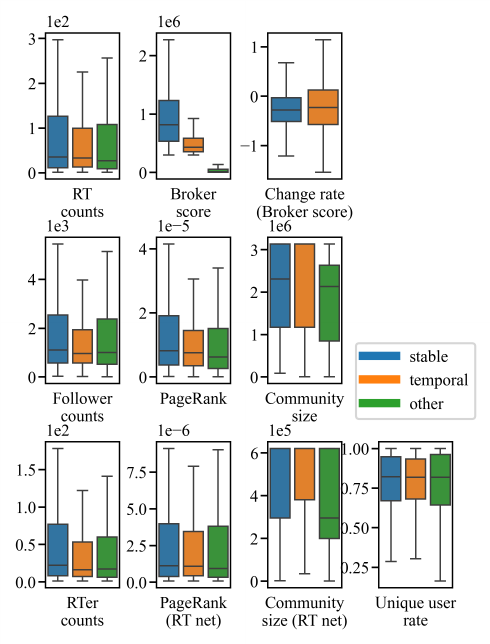}
    \label{fig1-3}
    }
\caption{Comparison of features among stable and temporal influencers and other users. (a) For source spreaders, distribution of RT counts, RTer counts, PageRank scores, and unique user rate show notable differences. (b) For brokers, distribution of broker scores shows a notable difference between groups. No substantial differences are observed in other feature distributions.}
\label{fig1-2fig1-3}
\end{figure*}

We now address \textbf{(RQ1)} by examining how the structural and behavioral characteristics of users differ across stable influencers, temporal influencers, and other users. To this end, we compare the distributions of these features for each group as of January 2022. Figure~\ref{fig1-2fig1-3} present boxplots of these comparisons, with Fig.~\ref{fig1-2} showing the results for source spreaders and Fig.~\ref{fig1-3} for brokers.

From Fig.~\ref{fig1-2}, we observe a notable difference in the distributions of RT counts among source spreaders, with stable source spreaders tending to have higher RT counts. Similarly, Fig.~\ref{fig1-3} shows a significant difference in the distributions of broker scores among brokers, with stable brokers tending to have higher broker scores. Thus, for both types of influencers, the distributions of influencer scores—which define user influence—differ substantially between stable and other users. Conversely, RT counts among brokers and broker scores among source spreaders did not show notable differences.

Regarding other features, for source spreaders, differences in distributions were observed for RTer counts and PageRank scores derived from the RT network, with stable users tending to have higher values. Additionally, differences were found in the unique user rate, where stable users tended to have lower rates. Centrality metrics derived from the follower network showed smaller median differences, but the overall distributions tended to be higher for stable users. In contrast, for brokers, these centrality metrics and features did not exhibit notable differences in distributions compared to source spreaders.

\subsection{Classification Models for Source Spreaders and Brokers}

We next address \textbf{(RQ2)} by evaluating the classification accuracy of our models for identifying stable influencers. Before analyzing the performance of machine learning–based models, we first establish a baseline using a simple influencer score. In this baseline, users are classified solely based on their current influence, with those holding the highest scores selected as stable influencers.
For evaluation, we matched the number of predicted stable influencers to the ground truth: 965 for source spreaders and 807 for brokers. Accordingly, the top 965 source spreaders and top 807 brokers ranked by influencer score were classified as stable influencers.
Table~\ref{tabbase} presents the classification results for this baseline model.
In the subsequent experiments, we use this baseline as a reference to assess the performance gains achieved by combining multiple features through machine learning. Note that despite its simplicity, the baseline model performs substantially better than random classification.

\begin{table}[tbp]
\renewcommand{\arraystretch}{1.1}
\caption{Classification Results for the Baseline Model}
\begin{center}
\begin{tabular}{l|r|r}
\hline
\textbf{Input features} & \textbf{\textit{AUC}}& \textbf{\textit{Accuracy}} \\
\hline
Source Spreader (Baseline)& 0.7673 & 0.6992\\
Broker (Baseline)& 0.8039 & 0.7414 \\
\hline
\end{tabular}
\label{tabbase}
\end{center}
\end{table}

Tables~\ref{tabSP} and \ref{tabBR} present the classification accuracies of our LightGBM models for predicting stable source spreaders and stable brokers, respectively. For each prediction task, we constructed four models: all Features Model, which uses all features listed in Tab.~\ref{tabF}, Follow Features Model, which uses only features the features categorized as "Follow", RT Features Model, which uses only features the features categorized as "RT" and Influence Score only Model, which uses only RT counts for source spreader prediction and broker scores for broker prediction. 

\begin{table}[tbp]
\renewcommand{\arraystretch}{1.1}
\caption{Classification performance for source spreaders}
\begin{center}
\begin{tabular}{l|r|r}
\hline
\textbf{Input features} & \textbf{\textit{AUC}}& \textbf{\textit{Accuracy}} \\
\hline
All features& 0.8895 & 0.8016\\
Follow features& 0.6207 & 0.6005 \\
RT features & 0.8844 & 0.8005\\
RT counts only & 0.8821 & 0.8016\\
\hline
\end{tabular}
\label{tabSP}
\end{center}
\end{table}
    
\begin{table}[tbp]
\renewcommand{\arraystretch}{1.1}
\caption{Classification Performance for Brokers}
\begin{center}
\begin{tabular}{l|r|r}
\hline
\textbf{Input features} & \textbf{\textit{AUC}}& \textbf{\textit{Accuracy}} \\
\hline
All features& 0.8024 & 0.7272\\
Follow features& 0.5146 & 0.5045 \\
RT features& 0.5156 & 0.5140\\
Broker score only& 0.8148 & 0.7298 \\
\hline
\end{tabular}
\label{tabBR}
\end{center}
\end{table}

Table~\ref{tabSP} shows that the all features model identifies stable source spreaders with high predictive performance, achieving an AUC of 0.89 and accuracy above 0.80. This performance exceeds that of the simple baseline presented in Tab.~\ref{tabbase}, demonstrating the effectiveness of machine learning in capturing the characteristics of stable source spreaders.
However, both the RT Features model and the RT Counts model achieve comparable performance to the all Features model. This suggests that stable source spreaders can be effectively identified using retweet-related data from the past four months alone.

Table~\ref{tabBR} shows that the Broker Score only model achieves the highest prediction accuracy among the tested models, with an AUC of 0.81 and accuracy of 0.73. Although its performance is lower than that for the source spreader prediction task, the model still identifies stable brokers with reasonably high performance. However, compared to the simple baseline in Tab.~\ref{tabbase}, the improvement offered by machine learning models is marginal.
These results suggest that a single month’s broker score is sufficient to identify stable brokers, and that further improvements in prediction performance would likely require additional features not explored in this study.

\begin{figure}[tbp]
\centering
\subfloat[Source spreader classification]{
    \includegraphics[width=0.9\linewidth]{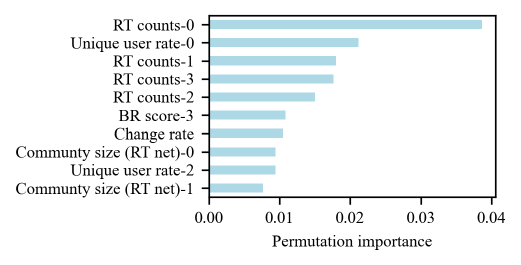}
    \label{fig2-1}
    }
\hfil
\subfloat[Broker classification]{
    \includegraphics[width=0.9\linewidth]{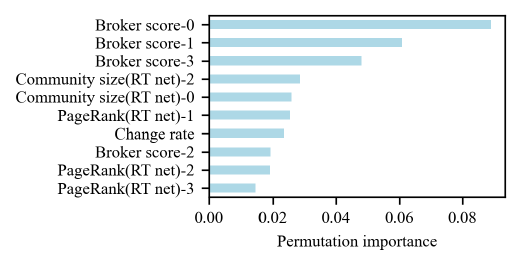}
    \label{fig3-1}
    }
\caption{Permutation importance of features used in the all features model for stable influencers prediction. (a) Top 10 important features for predicting stable source spreaders, with current RT counts, unique user rate, and past RT counts ranking highest. (b) Top 10 important features for predicting stable brokers, with current and past broker scores being the most influential.}
  \label{fig2-1fig3-1}
\end{figure}

Furthermore, we address \textbf{(RQ3)} by evaluating  the feature importance of our models for identifying stable influencers. 
Figure~\ref{fig2-1fig3-1} show the most important features in the models using all features for source spreaders and brokers, respectively.
To assess feature importance, we employ Permutation Importance~\cite{randomforest}. This method measures the decrease in model performance when the values of a given feature are randomly shuffled, thereby indicating how much the model relies on that feature for accurate predictions. 
The numbers following the feature names indicate how many months prior the data correspond to; 0 means the feature is from the reference month, January.

Figure~\ref{fig2-1fig3-1} show that influencer scores were found to be the most important features for both influencer types.
For source spreaders as shown in Fig.~\ref{fig2-1}, the current RT counts was the most important feature, followed by unique user rate and past RT counts. Follow-related features rarely appeared among the top important features and contributed little to classification.
For brokers as shown in Fig.~\ref{fig3-1}, broker-related features dominated in importance. The current broker score was the most important, with past broker scores also ranking highly.
These findings indicate that broker scores are critical for broker classification, while RT and follow information alone do not effectively contribute.

\begin{figure*}[!b]
\centering
\subfloat[Source spreader classification]{
    \includegraphics[width=0.45\linewidth]{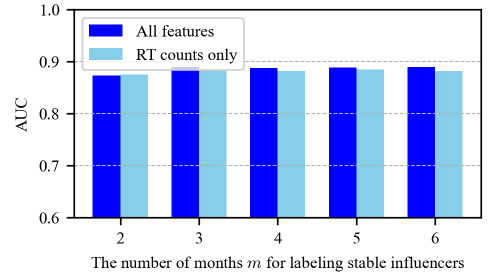}
    \label{fig2-X2}
    }
\hfil
\subfloat[Broker classification]{
    \includegraphics[width=0.45\linewidth]{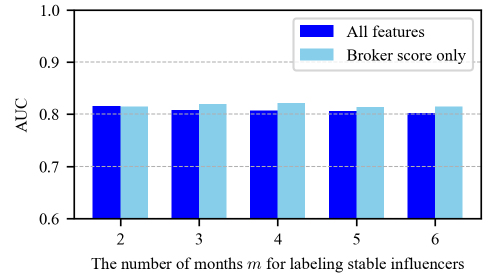}
    \label{fig3-X2}
    }
  \caption{AUC for classification with different labeling periods $m$. (a) The AUC stays around 0.9 for source spreaders. (b) The AUC stays around 0.8 for brokers.}
    \label{fig2-X2fig3-X2}
\end{figure*}

Next, we investigate how the classification performance changes when the period used to label stable influencers during model training is shortened from $m=6$.
From a practical standpoint, such as marketing applications, the required duration to build the model is an important factor.
So far, the models have used $m=6$, requiring a six-month period to assign training labels. Here, we examine the impact on model performance when $m$ is shortened.
Figure~\ref{fig2-X2fig3-X2} show the AUC scores for source spreaders and brokers, respectively, when $m$ varies from 2 to 6.
Each graph presents results using all features and using only the influencer score.
Note that users labeled as stable influencers in the test data are those who were consistently identified as influencers over a continuous six-month period.

From Figs.~\ref{fig2-X2fig3-X2}, we can find that the AUC for source spreader classification remains close to 0.9 regardless of $m$. Although there is a slight decrease at $m=2$, the overall classification performance does not significantly change. Similarly, for brokers, the AUC stays around 0.8 and varies little with $m$.
These results suggest that a long observation period is not necessary for model construction, and stable influencers can be labeled based on influence persistence of about 2 to 3 months without substantial loss of accuracy.

\begin{figure*}[!t]
\centering
\subfloat[Source spreader classification]{
    \includegraphics[width=0.45\linewidth]{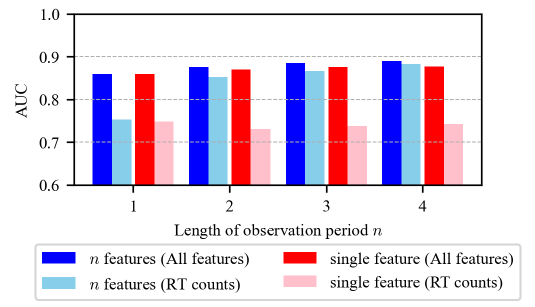}
    \label{fig2-X}
    }
\hfil
\subfloat[Broker classification]{
    \includegraphics[width=0.45\linewidth]{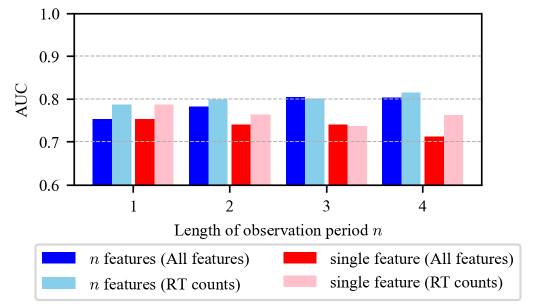}
    \label{fig3-X}
    }
\caption{AUC for influencer classification with different feature collection periods. (a) For source spreaders, AUC remains stable when using all features, regardless of the number of months. When using only the RT score, AUC improves markedly for $n \geq 2$, approaching that of the all features model. (b) For brokers,  the broker score only model achieves AUC comparable to or exceeding that of the model using all features.}
\label{fig2-Xfig3-X}
\end{figure*}

Finally, we analyze how classification performance varies depending on the duration of the feature collection period. In previous experiments, features were collected over four months for classification. From a practical perspective, particularly in marketing contexts, the length of the feature collection period is an important consideration. In this section, we vary the observation period 
$n$ from 1 to 4 months and observe the resulting changes in classification performance. Furthermore, while previous experiments extracted $n$ features from $n$ months of data, we also compare this with a method that calculates only a single feature from $n$ months of data. Figure~\ref{fig2-Xfig3-X} show the classification performance for identifying source spreaders and brokers, respectively.

As shown in Fig.~\ref{fig2-X}, when using all features for source spreader classification, AUC remains largely consistent across different feature collection periods and numbers of features. However, in models that rely solely on the RT score, AUC improves noticeably when $n\geq2$, reaching levels comparable to the model using all features. In contrast, aggregating the RT score into a single feature over a longer period does not produce similar benefits. This indicates that disaggregated temporal representations, such as multiple monthly RT features, are more effective for classification than a single cumulative metric. Furthermore, even with data from only one month, incorporating additional features such as the unique user ratio and change rate leads to substantially better performance than using RT counts alone.

Figure~\ref{fig3-X} indicates that increasing the number of features slightly improves broker classification performance, while performance decreases for single-feature models as the collection period extends. Notably, the model based solely on the broker score achieves performance comparable to or exceeding that of the model using all features. This implies that for broker classification, monitoring the broker score alone is more effective than relying on complex network-based features.

\section{Conclusion and Future Work}
\label{sec:conc}

In this paper, we analyzed the characteristics of stable and temporal influencers on Twitter using long-term data, and developed models to classify stable influencers.
Our analysis revealed that only about half of the influencers maintained their influence over a six-month period, indicating that being an influencer in a given month does not necessarily imply long-term influence. We also found that brokers could not be characterized by a single centrality metric based on network structure.
In the classification of stable influencers, both current and past retweet counts were important features for identifying source spreaders, while the broker score was important for classifying brokers. Additionally, for source spreaders, the unique user rate also contributed to prediction, which may reflect the presence of dedicated fans who repeatedly retweet the same user. However, information from the follower network had low importance for both types of influencers. This suggests that determining incentives based on follower count in influencer marketing does not necessarily align with the true characteristics of influence.

This study has several limitations, which also suggest directions for future research. First, there is no single agreed-upon definition of an influencer. While this study focused on two specific types, source spreaders and brokers, it remains unclear how the results might differ when alternative definitions or additional types of influencers are considered.
Second, our findings indicate that, beyond the broker score, we were unable to identify other effective features for detecting stable brokers. To improve prediction performance and to enable classification based on features other than the broker score, future research could explore diversifying predictive features by incorporating latent features using techniques such as node embedding, as well as linguistic features extracted from users’ posts or profile descriptions.
Third, to assess the generalizability of our findings, it is essential to evaluate them using a more diverse set of algorithms and datasets. In this study, we employed only a single machine learning algorithm; therefore, a more comprehensive evaluation using multiple algorithms is needed. Additionally, our study is based on a one-year Twitter dataset. Evaluations over longer time span, as well as validation of the proposed method on other social media platforms such as Instagram, Facebook and Bluesky, are also required.

\section*{Acknowledgments}
This work was supported by JSPS KAKENHI Grant No. JP25K03105


\begin{thebibliography}{10}
\providecommand{\url}[1]{#1}
\csname url@samestyle\endcsname
\providecommand{\newblock}{\relax}
\providecommand{\bibinfo}[2]{#2}
\providecommand{\BIBentrySTDinterwordspacing}{\spaceskip=0pt\relax}
\providecommand{\BIBentryALTinterwordstretchfactor}{4}
\providecommand{\BIBentryALTinterwordspacing}{\spaceskip=\fontdimen2\font plus
\BIBentryALTinterwordstretchfactor\fontdimen3\font minus \fontdimen4\font\relax}
\providecommand{\BIBforeignlanguage}[2]{{%
\expandafter\ifx\csname l@#1\endcsname\relax
\typeout{** WARNING: IEEEtran.bst: No hyphenation pattern has been}%
\typeout{** loaded for the language `#1'. Using the pattern for}%
\typeout{** the default language instead.}%
\else
\language=\csname l@#1\endcsname
\fi
#2}}
\providecommand{\BIBdecl}{\relax}
\BIBdecl

\bibitem{whiting2013people}
A.~Whiting and D.~Williams, ``Why people use social media: a uses and gratifications approach,'' \emph{Qualitative Market Research: an International Journal}, vol.~16, no.~4, pp. 362--369, 2013.

\bibitem{aillerie2018social}
K.~Aillerie and S.~McNicol, ``Are social networking sites information sources? informational purposes of high-school students in using snss,'' \emph{Journal of Librarianship and Information Science}, vol.~50, no.~1, pp. 103--114, 2018.

\bibitem{arafah2023digital}
B.~Arafah and M.~Hasyim, ``Digital literacy on current issues in social media: Social media as a source of information,'' \emph{Journal of Theoretical and Applied Information Technology}, vol. 101, no.~10, pp. 3943--3951, 2023.

\bibitem{leung2022online}
F.~F. Leung, F.~F. Gu, and R.~W. Palmatier, ``Online influencer marketing,'' \emph{Journal of the Academy of Marketing Science}, vol.~50, no.~2, pp. 226--251, 2022.

\bibitem{campbell2020more}
C.~Campbell and J.~R. Farrell, ``More than meets the eye: The functional components underlying influencer marketing,'' \emph{Business Horizons}, vol.~63, no.~4, pp. 469--479, 2020.

\bibitem{Uehara}
M.~Uehara and S.~Tsugawa, ``Analysis of the evolution of the influence of central nodes in a twitter social network,'' in \emph{Proceedings of 2019 IEEE 43rd Annual Computer Software and Applications Conference}, vol.~1, 2019, pp. 892--895.

\bibitem{bakshy2011everyone}
E.~Bakshy, J.~M. Hofman, W.~A. Mason, and D.~J. Watts, ``Everyone's an influencer: quantifying influence on twitter,'' in \emph{Proceedings of the Fourth ACM International Conference on Web Search and Data Mining}, 2011, pp. 65--74.

\bibitem{tsugawa2023identifying}
S.~Tsugawa and K.~Watabe, ``Identifying influential brokers on social media from social network structure,'' in \emph{Proceedings of the International AAAI Conference on Web and Social Media}, vol.~17, 2023, pp. 842--853.

\bibitem{BURT2000345}
R.~S. Burt, ``The network structure of social capital,'' \emph{Research in Organizational Behavior}, vol.~22, pp. 345--423, 2000.

\bibitem{Araujo04052017}
T.~Araujo, P.~Neijens, and R.~Vliegenthart, ``Getting the word out on twitter: the role of influentials, information brokers and strong ties in building word-of-mouth for brands,'' \emph{International Journal of Advertising}, vol.~36, no.~3, pp. 496--513, 2017.

\bibitem{pei2014searching}
S.~Pei, L.~Muchnik, J.~S. Andrade, Jr, Z.~Zheng, and H.~A. Makse, ``Searching for superspreaders of information in real-world social media,'' \emph{Scientific Reports}, vol.~4, no.~1, p. 5547, 2014.

\bibitem{LI20145115}
J.~Li, W.~Peng, T.~Li, T.~Sun, Q.~Li, and J.~Xu, ``Social network user influence sense-making and dynamics prediction,'' \emph{Expert Systems with Applications}, vol.~41, no.~11, pp. 5115--5124, 2014.

\bibitem{freeman1978centrality}
L.~C. Freeman, ``Centrality in social networks conceptual clarification,'' \emph{Social Networks}, vol.~1, no.~3, pp. 215--239, 1978.

\bibitem{LU20161}
L.~Lü, D.~Chen, X.-L. Ren, Q.-M. Zhang, Y.-C. Zhang, and T.~Zhou, ``Vital nodes identification in complex networks,'' \emph{Physics Reports}, vol. 650, pp. 1--63, 2016.

\bibitem{bucur2020top}
D.~Bucur, ``Top influencers can be identified universally by combining classical centralities,'' \emph{Scientific Reports}, vol.~10, no.~1, p. 20550, 2020.

\bibitem{kempe2003maximizing}
D.~Kempe, J.~Kleinberg, and {\'E}.~Tardos, ``Maximizing the spread of influence through a social network,'' in \emph{Proceedings of the ninth ACM SIGKDD International Conference on Knowledge Discovery and Data Mining}, 2003, pp. 137--146.

\bibitem{bhowmick2019temporal}
A.~K. Bhowmick, M.~Gueuning, J.-C. Delvenne, R.~Lambiotte, and B.~Mitra, ``Temporal sequence of retweets help to detect influential nodes in social networks,'' \emph{IEEE Transactions on Computational Social Systems}, vol.~6, no.~3, pp. 441--455, 2019.

\bibitem{huang2020identifying}
X.~Huang, D.~Chen, D.~Wang, and T.~Ren, ``Identifying influencers in social networks,'' \emph{Entropy}, vol.~22, no.~4, p. 450, 2020.

\bibitem{Pei2018}
S.~Pei, F.~Morone, and H.~A. Makse, \emph{Theories for Influencer Identification in Complex Networks}.\hskip 1em plus 0.5em minus 0.4em\relax Cham: Springer International Publishing, 2018, pp. 125--148.

\bibitem{cha2010measuring}
M.~Cha, H.~Haddadi, F.~Benevenuto, and K.~Gummadi, ``Measuring user influence in twitter: The million follower fallacy,'' in \emph{Proceedings of the International AAAI Conference on Web and Social Media}, vol.~4, no.~1, 2010, pp. 10--17.

\bibitem{ye2010measuring}
S.~Ye and S.~F. Wu, ``Measuring message propagation and social influence on twitter.com,'' in \emph{Proceedings of the Second International Conference on Social Informatics}.\hskip 1em plus 0.5em minus 0.4em\relax Springer, 2010, pp. 216--231.

\bibitem{PageRank}
S.~Brin and L.~Page, ``The anatomy of a large-scale hypertextual web search engine,'' \emph{Computer Networks and ISDN Systems}, vol.~30, no.~1, pp. 107--117, 1998.

\bibitem{Leiden}
V.~A. Traag, L.~Waltman, and N.~J. Van~Eck, ``From louvain to leiden: guaranteeing well-connected communities,'' \emph{Scientific Reports}, vol.~9, no.~1, pp. 1--12, 2019.

\bibitem{lightgbm}
G.~Ke, Q.~Meng, T.~Finley, T.~Wang, W.~Chen, W.~Ma, Q.~Ye, and T.-Y. Liu, ``{LightGBM}: A highly efficient gradient boosting decision tree,'' \emph{Advances in Neural Information Processing Systems}, vol.~30, 2017.

\bibitem{randomforest}
L.~Breiman, ``Random forests,'' \emph{Machine Learning}, vol.~45, pp. 5--32, 2001.

\end{thebibliography}


\end{document}